\documentclass{article}

\usepackage{amssymb,amsmath,epsfig,latexsym,graphicx,bm}

\usepackage[letterpaper,  margin=1in]{geometry}

\usepackage{float}
\usepackage{epstopdf}
\DeclareGraphicsExtensions{.eps}
\usepackage{multicol}
\usepackage{cite}
\usepackage{color}
\usepackage{soul}
\usepackage[ruled]{algorithm}
\usepackage{algcompatible}

\DeclareMathOperator*{\argmax}{\arg\!\max}
\algnewcommand\INPUT{\item[\textbf{Input:}]}%
\algnewcommand\OUTPUT{\item[\textbf{Output:}]}%

\usepackage{epstopdf}
\usepackage{subfigure}

\usepackage{setspace}

\usepackage{authblk}

\begin{document}

\date{ }
\title{SPECMAR: Fast Heart Rate Estimation from PPG Signal using a Modified Spectral Subtraction Scheme with Composite Motion Artifacts Reference Generation}

\author{Mohammad Tariqul Islam\thanks{mhdtariqul@gmail.com}}
\author{Sk. Tanvir Ahmed \thanks{tanvir@outlook.com}}
\author{Celia Shahnaz\thanks{celia@eee.buet.ac.bd}}
\author{Shaikh Anowarul Fattah\thanks{fattah@eee.buet.ac.bd}}
\affil{BUET, Dhaka, Bangladesh}

%





\maketitle

\begin{abstract}

The task of heart rate estimation using photoplethysmographic (PPG) signal is challenging due to the presence of various motion artifacts in the recorded signals. In this paper, a fast algorithm for heart rate estimation based on modified SPEctral subtraction scheme utilizing Composite Motion Artifacts Reference generation (SPECMAR) is proposed using two-channel PPG and three-axis accelerometer signals. First, the preliminary noise reduction is obtained by filtering unwanted frequency components from the recorded signals. Next, a composite motion artifacts reference generation method is developed to be employed in the proposed SPECMAR algorithm for motion artifacts reduction. The heart rate is then computed from the noise and motion artifacts reduced PPG signal. Finally, a heart rate tracking algorithm is proposed considering neighboring estimates. The performance of the SPECMAR algorithm has been tested on publicly available PPG database. The average heart rate estimation error is found to be 2.09 BPM on 23 recordings. The Pearson correlation is 0.9907. Due to low computational complexity, the method is faster than the comparing methods. The low estimation error, smooth and fast heart rate tracking makes SPECMAR an ideal choice to be implemented in wearable devices.


\end{abstract}

\mbox{}
\vfill
Accepted in Medical and Biological Engineering and Computing. This is a pre-copyedit version.\\

http://dx.doi.org/10.1007/s11517-018-1909-x\\

\section{Introduction}

Monitoring life signals using wearable devices have become a ubiquitous technology in recent years. One of the most common way to monitor life signals in such devices is by using photoplethysmographic (PPG) signals~\cite{jain2014heart}. The photoplethysmography is an electro-optic technique, where a {PPG sensor} is placed above skin and the light reflected from the tissues is captured by a photo detector~\cite{tamura2014wearable}. This reflected light varies with respect to the amount of blood present in the tissue which varies with cardiac cycle. This signal which varies as cardiac cycle is called the PPG signal. PPG signals have various usage for wearable electronics, which include measurement of heart rate~\cite{allen2007photoplethysmography}, arterial blood oxygen saturation~\cite{kyriacou2002investigation}, blood pressure~\cite{mccombie2006adaptive} and reducing false arrythmia alarm reduction in ICU~\cite{gambarotta2016review}. Traditionally, electrocardiogram (ECG) is used for monitoring the heart rate. However, ECG is hardly suitable for day to day usage for its cumbersome sensor placement in the body~\cite{islam2017cascade}. In this context, PPG is in a favorable position for its low cost and simple acquisition mechanism of the {PPG sensor}. However, depending on the mounting mechanism of the {sensor} the captured PPG signal can be heavily corrupted by motion artifacts which is generally considered as an impediment to estimation of heart rate~\cite{zhang2015troika}. In this section, the recent methods have been discussed and then the scope in the domain is discussed along with our contribution.

A number of methods have been developed to reduce motion artifacts in the PPG signal and to compute heart rate. Kim~\emph{et~al.}~\cite{kim2006motion} employed independent component analysis to remove motion artifacts in PPG signals. A Kalman filter based method is proposed in~\cite{seyedtabaii2008kalman} for adaptive reduction of motion artifacts from the PPG signals. Fu~\emph{et~al.}~\cite{fu2008heart} employed wavelet method to extract heart rate from PPG signal. In a normalized least mean square adaptive scheme a difference of two channel PPG signal was used as reference to denoise the PPG signals~\cite{yousefi2014motion}. Zhang~{et~al.}~\cite{zhang2015troika} introduced a large PPG dataset with ground truth to perform heart rate estimation from PPG signal during exercise, which inspired numerous researchers to develop and test different methods. Briefly, the TROIKA~\cite{zhang2015troika} and JOSS~\cite{zhang2015photoplethysmography} are the first two methods to report the results on part of the database. After that a number of researchers worked to solve the problem efficiently. In this context, a detailed review on the state-of-the-art of PPG based heart rate estimation techniques is presented in~\cite{tamura2018photoplethysmogram, periyasamy2017review}. Additionally, Temko~\cite{temko2015estimation} proposed a heart rate estimation method where the wiener filter and phase vocoder are employed to remove motion artifacts from the PPG signals. Instead of performing only sequential feed-forward operations to remove noise, Islam~\emph{et~al.}~\cite{islam2017time} removed the motion artifacts parallely and independently in time and frequency domain. Later, network of adaptive filters is explored in time domain using cascade and parallel combination (CPC) of adaptive filters to achieve faster and more accurate results~\cite{islam2017cascade}. One major concern in time domain filtering based approach is that, these methods require filtering in each time step thus causing considerable delay. As a result, a method that avoids such time domain computation yet provides less computational burden is still in great demand.

We present a spectral subtraction based motion artifacts removal scheme from PPG signals utilizing accelerometer data as motion artifacts reference. In this paper, a modified SPEctral subtraction method is proposed along with a Composite Motion Artifacts Reference generation, referred to as SPECMAR. First, the signals are pre-filtered to make them band limited and to remove unwanted frequency components. Next, the three channel accelerometer data are combined to produce a composite motion artifacts spectrum. A modified spectrum subtraction scheme is then introduced to obtain motion artifacts reduced PPG signal. The heart rate is computed from the noise reduced spectrum. Finally, a tracking algorithm is proposed to finalize heart rate estimate from the obtained spectrum. The proposed method has been tested on a publicly available database from~\cite{zhang2015troika} and has been compared with some state-of-the-art methods.

\section{Materials and Methods}

\subsection{Data and Task Specification} \label{sec:data}

The data used in this study was introduced by Zhang~\emph{et~al.} in the TROIKA framework \cite{zhang2015troika}, where results of only $12$ subjects were presented. {Each recording consists of two channel PPG signals, three-axis acceleration signal and single channel ECG signal sampled at 125 Hz for approximately 5 minutes.} Later, data of additional $11$ {recordings have been published where only the $13$th recording contains the ECG data. The two channel  PPG signals were recorded from wrist using a pulse oximeter with  green  LEDs of 515 nm wavelength which was used in reflection PPG setup. The distance from center to center of the PPG sensors was 2 cm. The three-axis acceleration signals were also and one channel chest ECG signal were recorded simultaneously. Both the pulse oximeters and the accelerometer were embedded in a wristband which was asked to be worn comfortably.} 
{During data recording subjects walked or ran on a treadmill with a speed of 1-2 km/hour for 0.5 minute, followed by a speed of 6-8 km/hour for 1 minute, a speed of 12-15 km/hour for 1 minute, a speed of 6-8 km/hour for 1 minutes, a speed of 12-15 km/hour for 1 minute, and a speed of 1-2 km/hour for 0.5 minute{~\cite{zhang2015troika}}. The subjects were asked to use the hand with the wristband to perform various tasks, such as, to pull clothes, wipe sweat on forehead, and push buttons on the treadmill, in addition to freely swing.}
The $13$th recording is of a subject {who was also for running on treadmill using the same exercise routine as of the first $12$ recording. However, the subject had abnormal heart rhythm.} In the recordings {$13$, $14$, $15$, $20$ and $23$}, the subjects performed various forearm and upper arm exercise (e.g. shake hands, stretch, push), running, jumping, or push-ups. {In the rest of the recordings, the subjects performed} intensive forearm and upper arm movements (e.g. boxing). {The 23rd recording is of a female subject with abnormal heart rhythm and blood pressure.}

{In addition to the signals each recordings contain ground truth heart rates computed from the ECG signal.} The ground truth heart rate has been computed from a $8$s time frame. Each time frame has an overlap of $6$s with the previous time frame.

\subsection{Proposed Method}
The proposed heart rate estimation method consists of three major stages. They are pre-processing of the signals, performing spectral subtraction using composite motion artifacts data to remove motion artifacts and estimation of heart rate. Each of the steps are described in the subsections that follow.

\subsubsection{Preprocessing}
The PPG signal has its biological origin that corresponds to the cardiac activity and thus widely used for heart rate measurements. Human heart rate generally varies between $0.5$ to $3.0$ Hz, even under intensive physical exercise. Thus all the PPG signals available are bandpass filtered from $0.4$ to $3.5$ Hz using a spectral domain filtering. {In the filtering scheme, the spectral coefficients of the fast Fourier transform (FFT) outside the range of $0.4$ to $3.5$ Hz are set to zero. Since, rest of the operations in the proposed method requires only the magnitude of the spectral coefficients within the range of the filtering, this is found to be a fast and efficient method for discarding the unwanted spectral coefficients. In this paper, the inverse fast Fourier transform (IFFT) is performed to show visualizations of the signal where only the real part of the IFFT is considered.} For the sake of consistency, a similar bandpass filtering is performed on each of the three axis accelerometer signals to cancel out the motion information beyond the specified frequency range. After the filtering operation, all the signals are normalized. Since the {PPG sensors} were generally placed very close, instead of considering the two acquired PPG signals separately, an average of the two signals is utilized as in~\cite{islam2017time}.

The spectra of the averaged PPG and three accelerometer signals are computed using the {FFT algorithm} with an order of $N_{FFT}$ {which is also the number of frequency bins in the spectrum}. In the magnitude spectrum of a frame of PPG data upto samples corresponding to $3.5$ Hz are considered and the rest of the samples are discarded. The truncated spectrum of the signals, which contain first $M$ FFT points corresponding to $H$ BPM, are termed as the pre-processed PPG spectrum ($\mathbf{X}_{PPG}$) and acceleration spectrum ($\mathbf{C}_X$, $\mathbf{C}_Y$, and $\mathbf{C}_Z$) are obtained. This action does not hamper the heart rate estimation due to the fact that no useful information related to heart rate lies beyond this specified range.
The value of $M$ is computed as
\begin{equation}
M =  \frac{H \times N_{FFT}}{60f_s}
\end{equation}
where $f_s$ is the sampling frequency.

\subsubsection{Proposed Modified Spectral Subtraction and Composite Motion Artifacts Reference Generation Scheme} 
The greatest advantage of spectral subtraction algorithm lies in its simplicity.
While using spectral subtraction, the underlying assumption is that noise is additive and stationary or slowly time varying signal, whose spectrum does not change significantly between
the updating periods. However, the main assumption of noise being stationary does not hold in PPG and sometimes the noise spectrum changes significantly between two successive epochs. In case of PPG, improvement of signal intelligibility is the main concern along with quality. Thus special cares have to be taken to implement spectral subtraction in PPG signal processing. In what follows, challenges and observations made while using the traditional spectral subtraction for PPG signals are presented followed by the considerations for generating composite motion artifacts spectrum, and finally the proposed modified spectral subtraction method is described.
 
\textbf{Generalized Spectral Subtraction:} By denoting the spectrum of noisy observed PPG signal as $\mathbf{Y}(k)$ and the noise spectrum as $\mathbf{N}(k)$, a generalized direct spectral subtraction algorithm performs the following operation

\begin{equation}
|\mathbf{X}(k)|^p= 
\begin{cases}
|\mathbf{Y}(k)|^p - |\mathbf{N}(k)|^p, \ \ \text{if} \ |\mathbf{Y}(k)|^p>|\mathbf{N}(k)|^p
\\
0, \ \ \ \ \ \ \ \ \ \ \ \ \ \ \text{otherwise}
\end{cases}
\label{Eq:SS_normal}
\end{equation} 

\begin{figure*}[t]
  \centering
  \includegraphics[scale=0.7]{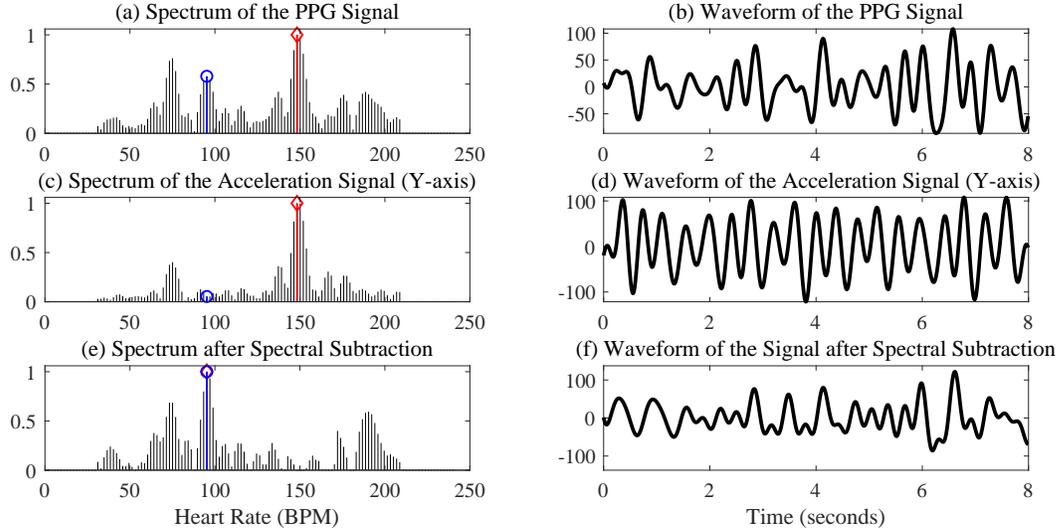}\\
  \caption{An example of Case I where motion artifacts and heart rate peaks are separable and motion artifacts peaks are also clearly visible in PPG signal. {The blue circle and red diamond correspond to location of true heart rate peak and dominant peak, respectively.}}\label{Fig:case1}
\end{figure*}

\begin{figure*}[t]
  \centering
  \includegraphics[scale=0.7]{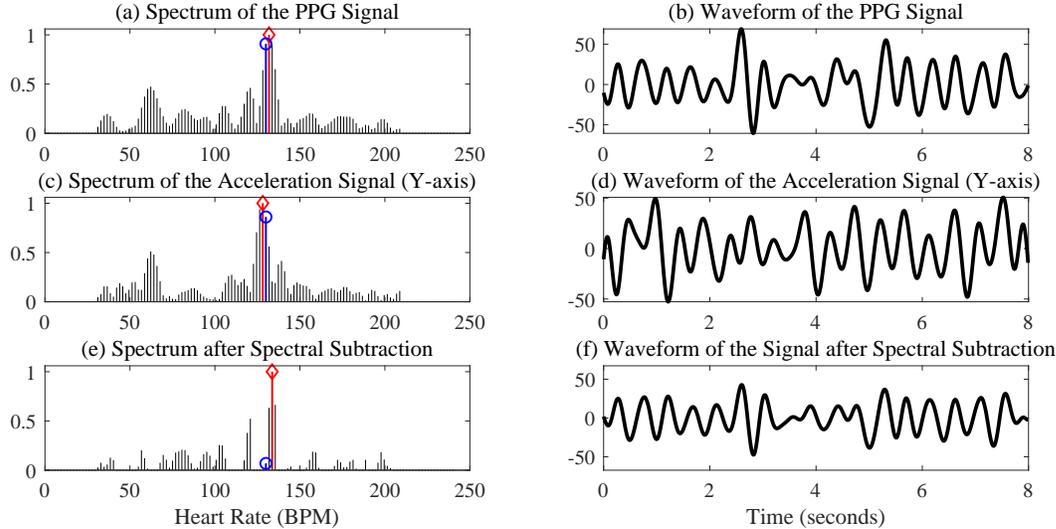}\\
  \caption{An example of Case II, where heart rate peaks and motion artifacts peaks are overlapping, thus non-separable by naked eye. However, after spectral subtraction a peak near to the true heart rate is dominant. {The blue circle and red diamond correspond to location of true heart rate peak and dominant peak, respectively.} }\label{Fig:case2}
\end{figure*}

\begin{figure*}[t]
  \centering
  \includegraphics[scale=0.7]{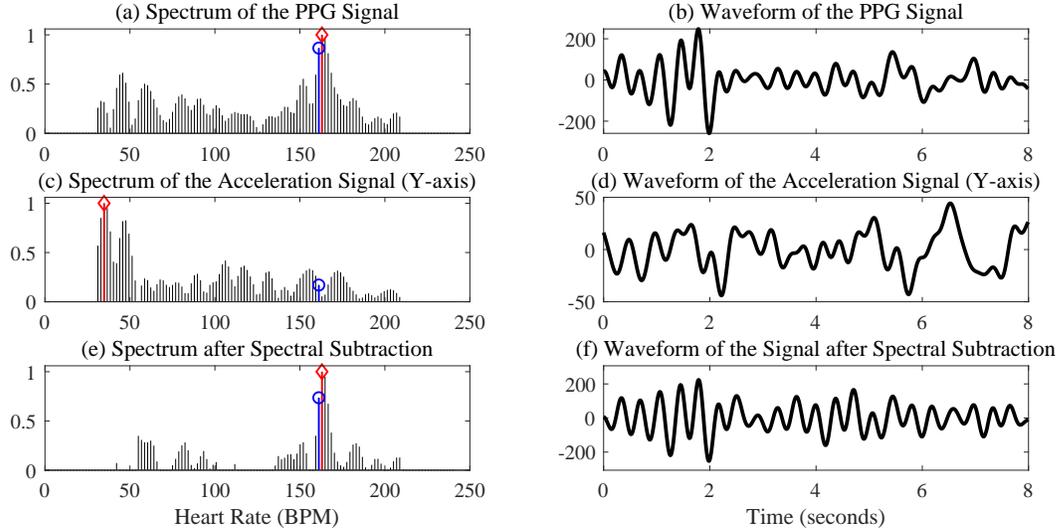}\\
  \caption{An example of Case III, where the acceleration signal does not bear much information about motion artifacts in the PPG signal. {The blue circle and red diamond correspond to location of true heart rate peak and dominant peak, respectively.}}\label{Fig:case3}
\end{figure*}

\begin{figure*}[t]
  \centering
  \includegraphics[scale=0.8]{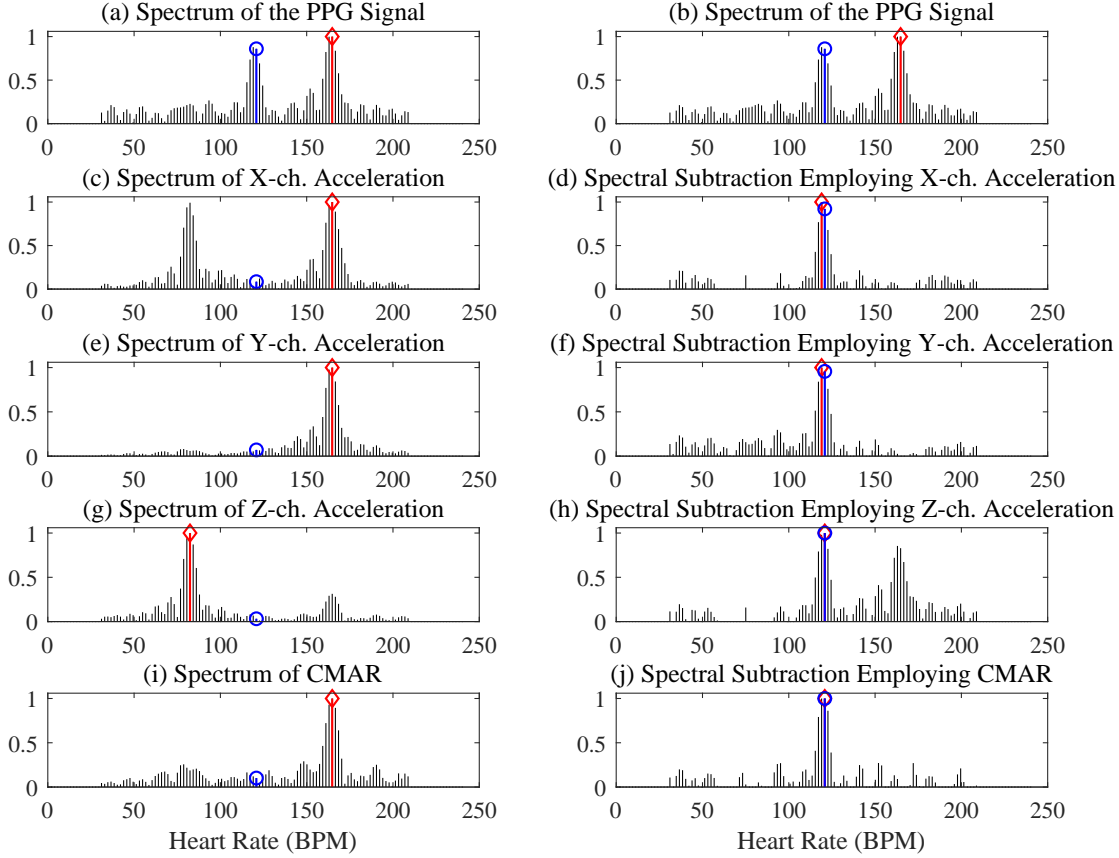}\\
  \caption{{Comparison of the spectra generated after spectral subtraction using different motion artifacts references. (a),(b) Spectrum of the PPG signal. Spectrum of the motion artifacts reference using (c) X-axis acceleration signal $\mathbf{C}_X$, (e) Y-axis acceleration signal $\mathbf{C}_Y$, (g) Z-axis acceleration signal $\mathbf{C}_Z$, and (i) using the proposed motion artifacts reference generation method $\mathbf{N}_{CMAR}$. (d), (f), (h), (j) The spectra generated after employing spectrum subtraction on the references (c), (e), (g), and (i), respectively. The blue circle and red diamond correspond to location of true heart rate peak and dominant peak, respectively.} }\label{Fig:ACC_Spectrum}
\end{figure*}

\begin{figure*}[t]
  \centering
  \includegraphics[scale=0.7]{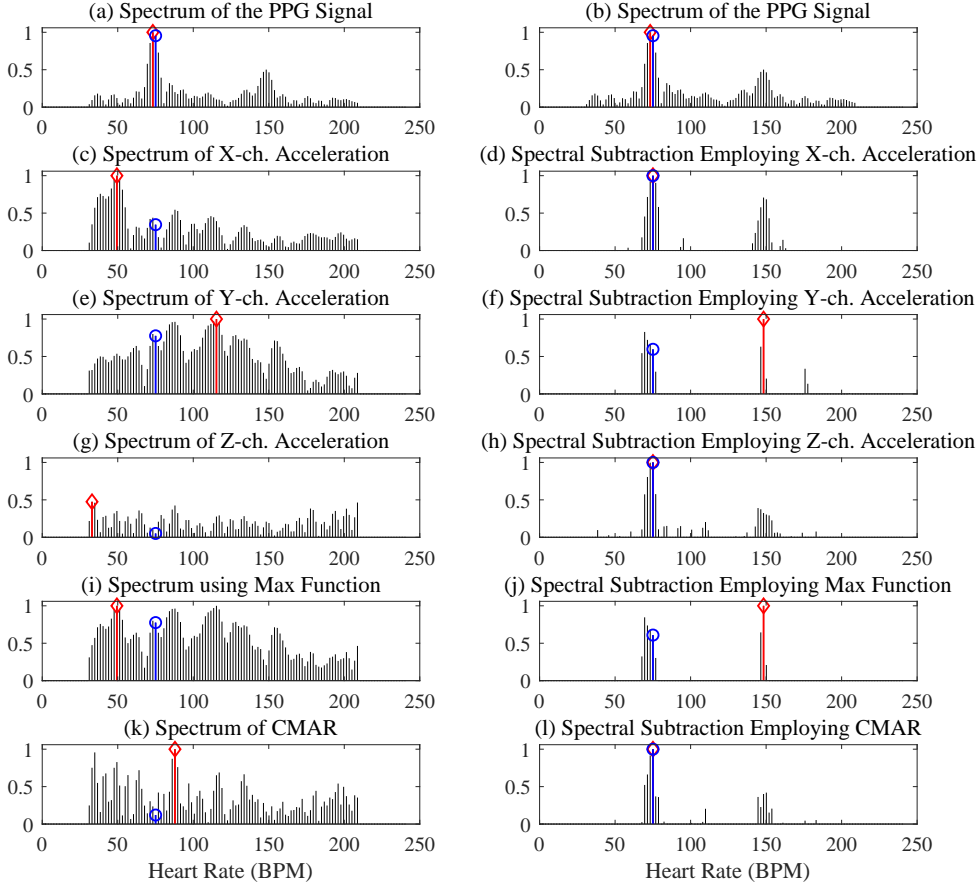}\\
  \caption{{Comparison of the spectra generated after spectral subtraction using different motion artifacts references. (a),(b) Spectrum of the PPG signal. Spectrum of the motion artifacts reference using (c) X-axis acceleration signal $\mathbf{C}_X$, (e) Y-axis acceleration signal $\mathbf{C}_Y$, (g) Z-axis acceleration signal $\mathbf{C}_Z$, (i) using the $\max(\mathbf{C}_X[k], \mathbf{C}_Y[k], \mathbf{C}_Z[k])$ fucntion and (k) using the proposed motion artifacts reference generation method $\mathbf{N}_{CMAR}$. (d), (f), (h), (j) and (l) The spectra generated after employing spectrum subtraction on the references (c), (e), (g), (i) and (k), respectively. The blue circle and red diamond correspond to location of true heart rate peak and dominant peak, respectively.} }\label{Fig:ACC_Spectrum_max}
\end{figure*}

\begin{figure*}[t]
  \centering
  \includegraphics[scale=0.8]{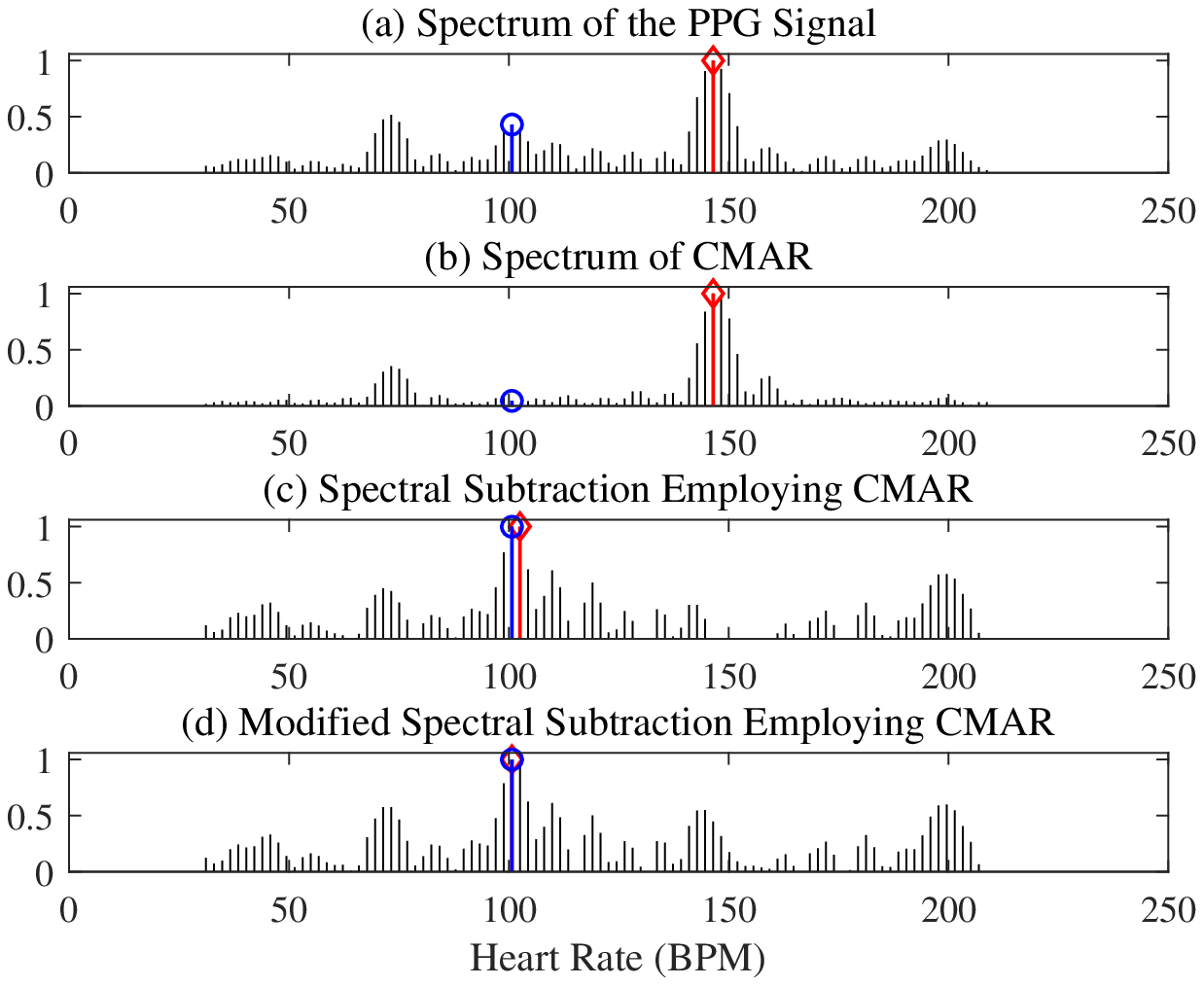}\\
  \caption{{Comparison of the spectra generated from spectral subtraction and proposed modified spectral subtraction using CMAR. (a) Spectrum of the PPG signal. (b) Spectrum of motion artifacts reference employing composite motion artifacts reference generation method. Spectrum obtained after (c) emloying spectral subtration without scaling and (d) employing modified spectral subtraction with $\alpha_1=0.88$ and $\alpha_2=0.70$. The blue circle and red diamond correspond to location of true heart rate peak and dominant peak, respectively.} }\label{Fig:MSS1}
\end{figure*}

where $\mathbf{X}$ is the spectrum of noise free signal, $p$ is the power exponent and $k=0,1,\ldots,M-1$ are the indexes of spectral components. Setting $p=1,2$ yields the magnitude and power spectral subtraction algorithm respectively. Throughout this paper the term spectral subtraction will refer to magnitude spectral subtraction. To mitigate the negative peak generation in basic spectral subtraction, a preset spectral floor is chosen when the noise spectrum component is higher than that in observation spectrum. This parameter has been set to zero in~(\ref{Eq:SS_normal}).

To demonstrate the performance of basic spectral subtraction and other variations of SS, we categorize PPG and acceleration spectrum in 3 broad categories-

Case I: (i) Motion artifacts peaks' locations exactly match with the peaks' locations of acceleration spectrum, (ii) true heart rate peak and motion artifacts peaks locate at separable distance in PPG spectrum.

Case II: (i) Motion artifacts peaks' locations exactly match with the peaks' locations of acceleration spectrum, (ii) motion artifacts peaks are in very close neighborhood of true heart rate peak or they overlap.

Case III: Acceleration signal does not bear any information about motion artifacts i.e. generate random peaks at random positions.

These three cases are investigated and some sample frames are depicted in Figs. \ref{Fig:case1}, \ref{Fig:case2} and \ref{Fig:case3}. The peaks corresponding to true heart rate and maximum peak in the spectra are presented using blue {circle and red diamond}, respectively. {In addition to spectra, the temporal PPG, the acceleration signal and the signal after spectral subtraction are provided to understad the morphology of the PPG signals before and after spectral subtraction. In order to construct the signal after spectral subtraction the phase information of the PPG signal is used.}  In time frames following Case I, where the motion artifacts peaks' locations in the PPG spectrum exactly match with the peaks' location of acceleration spectrum and they locate at distant neighborhood of heart rate peak, calculating difference spectrum by direct spectral subtraction of (\ref{Eq:SS_normal}) will generate a cleansed PPG spectrum. An example of such case is shown in fig. \ref{Fig:case1}. In this case, an acceleration signal ($\mathbf{C}_Y$) is chosen as motion artifacts reference signal. It is obvious that, the motion artifacts dominant peaks present in the acceleration spectrum, are also dominant in the pre-processed PPG spectrum and the true heart rate peak is dominant and at a separable distance. Thus after the basic spectral subtraction, the motion artifacts peaks are significantly reduced and only dominant peak corresponds to true heart rate. {It is seen from the waveforms that the high frequency oscillations corresponding to the motion artifacts are significantly reduced in the PPG signal after spectral subtraction.}

For cases II and III, the spectral subtraction of one or all of the acceleration signals from the pre-processed PPG signal cannot always recover the true heart rate peak. The examples of these cases are shown in Figs. \ref{Fig:case2} and \ref{Fig:case3}. In Case II shown in Fig.~\ref{Fig:case2}, though the dominant peak in the PPG represent the true heart rate, the motion artifacts peak in acceleration spectrum are at very close vicinity of the true heart rate peak (the movement rate of hand closely matches the heart rate). Thus performing basic spectral subtraction, the magnitude of the true heart rate peak is reduced and the dominant peak in the spectrum no longer corresponds to true heart rate. {Since, heart rate and motion artifacts peaks overlap, the waveforms of the PPG signal before and after spectral subtraction remain mostly unchanged.} In Fig.~\ref{Fig:case3}, the acceleration signal does not represent any motion artifacts. By subtracting it from the PPG spectrum, true heart rate peak location cannot be recovered. {It is observed from the waveforms of the signal that the morphology of the obtained signal after spectral subtraction is similar to that of the original PPG signal.}

However from these figures, it is observable that the true heart rate peak can be extracted by altering the basic spectral subtraction algorithm in (\ref{Eq:SS_normal}) in such a way that motion artifacts removal is facilitated. In order to improve the spectral subtraction for these above mentioned cases, beside modification to the actual algorithm, a careful choice of motion artifacts spectrum is also required. Motion artifacts spectrum can be constructed as one of the acceleration signals or a composite spectrum constructed from the three channel acceleration spectra. Such propositions are discussed in the next subsection.

\textbf{Motion Artifact Spectrum Generation:} 
Though motion at a certain time can be differently captured by any of the acceleration signals, a careful observation shows that, in most time frames, the motion artifacts peaks are available simultaneously in all acceleration signals. This is largely due to the periodic movement of hands during the physical activity and such movement may not necessarily be aligned in any particular direction of the accelerometer. For example, during running the hand movement appears as back and forth if the subject is seen from a side, appears as up and down if the subject is observed from behind and circular if seen from top. As a result, all of the acceleration signals capture similar frequencies of motion. Hence, by using only one of the acceleration signals in spectral subtraction may result in partial removal of motion artifacts from the PPG signal. On the other hand, using all three acceleration signals successively may result in diminishing the PPG signal. Hence, a carefully designed composite of three axis acceleration signal may provide satisfactory or desired performance.

It has already been discussed that, the channels of accelerometer experience movements with same period during a physical exercise. Since, all three acceleration signals are expected to have same frequency components of motion artifacts of different amplitude, the question thus arise is, for each frequency component which one to choose among the three components. Naively, the largest component for a frequency bin of the three spectra may be considered. However, sometimes the large component in one of the channels is not due to motion artifacts but due to noise in accelerometer. Since, motion artifacts is present in all three channels, choosing the minimum magnitude of the three spectra for a frequency bin captures the motion artifacts most efficiently. Thus, for the $k$th frequency bin of composite motion artifact reference {(CMAR)} spectrum $N_{CMAR}$ is formed by,
\begin{equation}
\mathbf{N}_{CMAR}(k) = \min(\mathbf{C}_X(k), \mathbf{C}_Y(k), \mathbf{C}_Z(k)). \label{Eq:ACC}
\end{equation}

{The composite spectrum is normalized in range $[0,1]$. 

Fig.{~\ref{Fig:ACC_Spectrum}} shows the effect of employing proposed composite motion artifacts reference generation method in spectrum subtraction. It is observed from the figure that, the accelerometer signals capture motion artifacts in similar frequencies. $\mathbf{C}_X$ captures motion artifact with approximately equal dominance in two frequencies, one fundamental and another at its double frequency. $\mathbf{C}_Y$ and $\mathbf{C}_Z$ capture these frequencies as well, however, only one of them is dominant. It is also seen from the figure that, the dominant motion artifact in the PPG signal includes only the higher frequency of the accelerometer signal. The composite motion artifact reference spectrum constructed by following{~(\ref{Eq:ACC})} is dominated by the higher frequency from the motion artifacts whereas the lower frequency motion artifact component is not dominant. Thus, in the spectrum after spectral subtraction, none of the motion artifacts frequencies are dominant and the location of the true heart rate and dominant peak coincide. In a similar way, Fig.{~\ref{Fig:ACC_Spectrum_max}} shows a case where the acceleration signals are noisy and does not correspond to motion artifact. In addition to previously considered references an alternative to ({\ref{Eq:ACC}}) is considered where instead of $\min(\cdot)$ function, the $\max(\cdot)$ function is considered for motion artifacts reference. In this case, the large components in the spectra of the acceleration signals are due to noise of the accelerometer and do not correspond to the motion artifacts. It is seen from the figure that after spectral subtraction, the $\max(\cdot)$ function causes degradation in the peak near the heart rate. On the other hand, using the proposed CMAR in the spectral subtraction, the dominant peak and the location of the heart rate coincide. Thus, the proposed CMAR can be an efficient motion artifacts reference for spectral subtraction denoising mechanism for PPG signals.}

\textbf{Modified Spectral Subtraction (MSS) Algorithm:}
In order for the spectral subtraction to work better with PPG, a modification of the basic spectral subtraction is required. Direct subtraction using~(\ref{Eq:SS_normal}) sometimes produces unwanted results, as seen in Figs. \ref{Fig:case2} and \ref{Fig:case3}. In formulation of~(\ref{Eq:SS_normal}), $|\mathbf{Y}(k)|$ is assumed to be the sum of $|\mathbf{X}(k)|$ and $|\mathbf{N}(k)|$. However, true $|\mathbf{N}(k)|$ is not available and a scaled version of $|\mathbf{N}(k)|$ is estimated as the motion artifacts spectrum. Thus we can replace {$|\mathbf{N}(k)|$} with $[\alpha_N |\mathbf{N}(k)|]$, where $\alpha_N$ is the scaling factor of the noise. Moreover, the $|\mathbf{Y}(k)|$ which is found from the PPG sensor is also an estimated or the scaled version of the internal sum of $|\mathbf{X}(k)|$ and $\alpha_N|\mathbf{N}(k)|$. Thus the expression of $\mathbf{Y}(k)$ is given by
\begin{equation}
|\mathbf{Y}(k)|=\alpha_Y (|\mathbf{X}(k)|+\alpha_N |\mathbf{N}(k)|), 
\end{equation}
where $\alpha_Y$ is the scaling factor of the observed output. Hence, we can write the estimation of $|\mathbf{X}(k)|$ as 
\begin{equation}
|\mathbf{X}(k)|=\frac{1}{\alpha_Y} |\mathbf{Y}(k)| - \frac{\alpha_N}{\alpha_Y} |\mathbf{N}(k)|.
\end{equation}

From the discussion above, the modified spectral subtraction (MSS) is proposed as
\begin{equation} 
|\mathbf{X(k)}|^p= 
\begin{cases}
\alpha_1|\mathbf{Y}(k)|^p - \alpha_2|\mathbf{N(k)}|^p, \ \ \text{if} \ \alpha_1|\mathbf{Y(k)}|^p>\alpha_2|\mathbf{N(k)}|^p
\\
0 \ \ \ \ \ \ \ \ \ \ \ \ \ \ \text{otherwise},
\end{cases}
\label{Eq:SS_prop}
\end{equation} 
where, $\alpha_1$ and $\alpha_2$ essentially works as the weight parameters. As both PPG signal spectrum ($\mathbf{X}=\mathbf{X}_{PPG}$) and noise signal spectrum ($\mathbf{N}=\mathbf{N}_{CMAR}$) values are in the range of $[0,1]$, the parameters, $\alpha_1$ and $\alpha_2$, can also be chosen in the range of $[0,1]$. As a result, the percentage of PPG signal and the percentage of noise spectrum can be logically adjusted. Controlling the amount of noise subtraction is more logical in handling PPG data, where no one can guarantee the actual amount of noise inserted in recorded PPG and thus expected proposed MSS is expected to provide better heart rate estimation. {Fig.{~\ref{Fig:MSS1}} shows the effect of the proposed modified spectrum subtraction. It is seen from the figure that, the dominant peak and the frequency bin corresponding to heart rate coincide when modified spectral subtraction is employed as opposed to the generalized spectral subtraction.} For convenience, the method employing this proposed modified SPEctral subtraction and Composite Motion Artifacts Reference generation is termed as SPECMAR. Note that, considering the values $\alpha_1=\alpha_2=1$ results in $\alpha_Y=1$ and $\alpha_N=1$, which is the case considered in (\ref{Eq:SS_normal}) with $\mathbf{N}=\mathbf{N}_{CMAR}$, is termed as SPECMAR without scaling (SPECMARWS).

\subsubsection{Heart Rate Estimation and Tracking}

Once the result of an individual frame is available, it is more logical to consider the results obtained in nearby frames and by incorporating an efficient tracking algorithm for estimating heart rate. The tracking algorithm proposed here consists of two main parts: initial estimation of heart rate at the beginning of the method and estimation and tracking of current heart rate.

\textbf{Initial Estimate:}
The method requires an initial estimate for the first time frame. The initial estimation is performed using the output of the modified spectral subtraction ($\mathbf{X}_{MSS}$). The highest peak of the spectrum is selected as the initial estimate. This estimated heart rate is used as the heart rate of the previous time frame ($N_0$) for the next frame.

\textbf{Estimation and Tracking of Heart Rate Using Best Candidate Approach:}
Even though the noise suppression using modified spectral subtraction is highly effective, there are some frames where it cannot resolve the heart rate peak. Most of them are due to motion artifacts components being in close vicinity of the heart rate component. In such cases, an error in heart rate estimation may occur. A proper estimation technique has to be employed to mitigate such error. One key point in estimating the heart rate successfully is that, the change of heart rate is generally not abrupt, rather gradual. If the time overlapping between two successive time frames is small, the true heart rate for latest frame will be close to that of the previous frame. Considering this, both pre-processed PPG spectrum ($\mathbf{X}_{PPG}$) and processed PPG spectrum ($\mathbf{X}_{MSS}$) are employed to track BPM correctly. The candidate peaks are selected from the $\mathbf{X}_{PPG}$ signal considering the search region, $\mathbf{P_0}=[N_0-\Delta_s, . . ., N_0+\Delta_s]$. That is, the neighborhood, $\pm\Delta_s$ bins, of the frequency bin of the previous time frame ($N_0$) are selected to be the most likely location of the heart rate component. Then peaks and their locations inside this search region are extracted. The location of the peaks that have magnitude greater than $25\%$ of the magnitude of maximum peak of $\mathbf{X}_{PPG}$ are put in the set $\mathbf{S_0}=[N_1,N_2,...]$. Now selection of location of heart rate component can be considered for three cases,

Case I: If only one peak is found in $\mathbf{S_0}$, that is cardinality of the set is $1$, this is the dominant peak, $D$ of the frame. Then in the close vicinity, $\pm\delta_1$ frequency bins, of the location of that peak, a search is performed in $\mathbf{X}_{MSS}$ to know whether that peak is present or a different peak has surfaced due to the motion artifact removal. The peak with highest magnitude in this search region is selected as the location of the heart rate peak ($N_{cur}$).

Case II: If more than one peak is found, it is suspected that motion artifacts or some other kind of noise is present in this spectrum and so emphasis is given on the dominant peak of the $\mathbf{X}_{PPG}$. If the dominant peak is within the range, $\pm\Delta_t$ frequency bins, of $N_0$, a search is performed in $\mathbf{X}_{MSS}$ within the $\pm\delta_2$ range of $N_0$ to find a satisfactory peak, that is the magnitude of the peak is at least $10\%$ of the dominant peak. The extracted location is set to $N_{cur}$.

Case III: If a satisfactory peak is not found in the two cases above, then all the peaks and their location in the search region $\mathbf{P_0}$ is extracted from $\mathbf{X}_{MSS}$. The set containing the normalized magnitude of the peaks is $\mathbf{A}$ and the set containing the location of the peaks is $\mathbf{S_0}$. Then a closeness to the heart rate location of previous frame is stored in set $\mathbf{C}$ as
\begin{equation}
\mathbf{C} = \left(1 - \frac{|\mathbf{S_0}-N_0|}{\kappa N_0} \right) ,\label{Eq:closeness}
\end{equation}
where $\kappa$ is a normalizing parameter chosen such that the values inside $\mathbf{C}$ is in the range $[0,1]$. Finally, the maximum value of weighted average of the peak amplitude set $\mathbf{A}$ and the closeness set $\mathbf{C}$ provides the estimated location of heart rate
\begin{equation}
i = \argmax(\beta_1 \mathbf{A} + \beta_2 \mathbf{C}) \label{Eq:calci}
\end{equation}
and the $i$th member of $\mathbf{S_0}$ is set to $N_{cur}$. Here, $\beta_1$ and $\beta_2$ are positive weight parameters satisfying $\beta_1+\beta_2=1$.

After that, heart rate in BPM is calculated to be
\begin{equation}
 \hat{B}_{est} = \frac{N_{cur}-1}{N_{FFT}}  \times 60  \times f_s. \label{Eq:Best1}
\end{equation}

Next in order to obtain a smooth heart rate tracking, a three point moving average is employed where the estimate of BPM in current frame is computed as
\begin{equation}
B_{est}' = \gamma_1 \hat{B}_{est} + \gamma_2 B_{-1} + \gamma_3 B_{-2}, \label{Eq:Best2}
\end{equation}
where $B_{-1}$ and $B_{-2}$ represent estimated BPM of previous two successive time frames and sum of the weight parameters satisfies $\gamma_1+ \gamma_2 + \gamma_3 = 1$ and these values are empirically chosen.   

Finally  in order to prevent extremely high or low estimated values of BPM in comparison to previous BPM estimate moving averaged value $B_{est}'$ is modified as
\begin{equation}
B_{est}= 
\begin{cases}
B_{est}'+5,\ \text{if} \ B_{est}'-B_{-1} \geq  \lambda_{inc}
\\
B_{est}'-3,\ \text{if} \ B_{est}'-B_{-1} \leq  \lambda_{dec}
\\
B_{est}',\ \ \ \ \ \ \text{otherwise},
\end{cases} \label{Eq:Best3}
\end{equation} 
where constants $\lambda_{inc}$ and $\lambda_{dec}$ are set empirically to small integer values.

\subsection{Parameter Settings}
The proposed method requires a number of values to be initialized. Most of these parameters are chosen empirically with reference to previous literature. The number of points to generate spectrum, $N_{FFT}$ is chosen to be $4096$. In order to produce truncated spectrum, $H$ is set to $240$. The weight parameters $\alpha_1$ and $\alpha_2$ of modified spectral subtraction are set to $0.88$ and $0.70$, respectively and the exponent $p$ is set to $1$. For large searching neighborhood, the parameters $\Delta_s$ and $\Delta_t$ are both set to $30$, whereas small search regions $\delta_1$ and $\delta_2$ are both set to $3$. The weight paramers $\beta_1$ and $\beta_2$ of~(\ref{Eq:calci}) are empirically set to $0.7$ and $0.3$. The weights $\gamma_1$, $\gamma_2$ and $\gamma_3$, of the moving average operation are set to $0.9$, $0.05$ and $0.05$, respectively and the parameters $\lambda_{inc}$ and $\lambda_{dec}$ are set to $5$ and $-3$, respectively as in~\cite{islam2017time} considering the rate of change of heart rate.

\begin{table*}
	\centering
	\caption{Performance comparison in terms of AAE for $23$ Data Recordings.} \label{Tab:AAE}
	\begin{tabular} {c c c c c c c c c c}
	\hline
	Data 					& TROIKA~\cite{zhang2015troika}	& MISPT-125~\cite{lakshminarasimha2015multiple} 	& WFPV~\cite{temko2015estimation} 	& TFD~\cite{islam2017time} 	& CPC~\cite{islam2017cascade} 	& SPECMARWS 	& SPECMAR\\
	\hline
	1 						& $2.29$ 			& $1.58$ 			& $1.23$ 			& $\mathbf{1.16}$	& $1.56$ 			& $1.22$ 			& $1.22$ \\
	2 						& $2.19$ 			& $1.80$ 			& $1.26$		 	& $\mathbf{1.16}$	& $2.25$ 			& $1.82$ 			& $1.51$ \\
	3 						& $2.00$ 			& $\mathbf{0.58}$ 	& $0.72$ 			& $0.79$ 			& $0.66$ 			& $0.78$ 			& $0.75$ \\
	4 						& $2.15$ 			& $0.99$ 			& $0.98$ 			& $0.87$ 			& $\mathbf{0.69}$	& $1.26$ 			& $1.26$\\
	5 						& $2.01$ 			& $0.74$  			& $0.75$ 			& $0.79$ 			& $\mathbf{0.71}$	& $0.83$ 			& $0.75$\\
	6 						& $2.76$ 			& $0.93$ 			& $\mathbf{0.91}$ 	& $1.14$ 			& $0.96$ 			& $1.88$ 			& $1.87$\\
	7 						& $1.67$ 			& $0.73$ 			& $\mathbf{0.67}$ 	& $0.71$ 			& $0.80$ 			& $1.06$ 			& $0.80$\\
	8 						& $1.93$ 			& $\mathbf{0.45}$  	& $0.91$ 			& $0.73$ 			& $0.56$ 			& $1.09$ 			& $1.07$\\
	9 						& $1.86$ 			& $\mathbf{0.41}$ 	& $0.54$ 			& $0.64$ 			& $0.56$ 			& $0.71$ 			& $0.65$\\
	10 						& $4.70$ 			& $3.60$ 			& $2.61$ 			& $3.09$ 			& $2.62$ 			& $3.48$ 			& $\mathbf{2.24}$\\
	11 						& $1.72$ 			& $\mathbf{0.88}$	& $0.94$ 			& $1.34$ 			& $1.27$ 			& $1.46$ 			& $1.39$\\
	12 						& $2.84$ 			& $\mathbf{0.68}$ 	& $0.98$ 			& $1.54$ 			& $0.79$ 			& $1.46$ 			& $1.09$\\
	\hline
	\textbf{Mean 12}  	& $\mathbf{2.34}$ 	& $\mathbf{1.11}$ 	& $\mathbf{1.05}$ 	& $\mathbf{1.16}$ 	& $\mathbf{1.12}$ 	& $\mathbf{1.33}$ 	& $\mathbf{1.21}$ \\
	\textbf{Std 12}  	& $\mathbf{2.47}$  	& $\mathbf{2.33}$  	& $\mathbf{1.33}$ 	& $\mathbf{1.74}$ 	& $\mathbf{2.30}$ 	& $\mathbf{1.77}$ 	& $\mathbf{1.75}$ \\
	\hline
	13 						& - 				& - 				& $3.58$ 			& $4.20$			& $\mathbf{3.49}$ 	& $3.98$ 			& $3.98$ \\
	14 						& - 				& - 				& $9.66$ 			& $6.69$			& $14.58$ 			& $6.83$ 			& $\mathbf{6.57}$\\
	15 						& - 				& - 				& $2.31$ 			& $2.29$			& $1.86$ 			& $2.00$ 			& $\mathbf{1.76}$\\
	16						& - 				& - 				& $4.93$ 			& $16.24$ 			& $10.19$ 			& $3.20$ 			& $\mathbf{2.28}$\\
	17 						& - 				& - 				& $3.07$ 			& $3.02$			& $7.62$ 			& $2.81$ 			& $\mathbf{2.77}$\\
	18 						& - 				& - 				& $2.67$ 			& $2.48$			& $\mathbf{1.30}$	& $3.18$ 			& $2.94$\\
	19 						& - 				& - 				& $\mathbf{3.11}$ 	& $4.27$			& $8.39$ 			& $5.53$ 			& $4.80$\\
	20 						& -					& - 				& $\mathbf{2.10}$ 	& $2.87$			& $2.53$ 			& $2.56$ 			& $2.72$\\
	21 						& -					& - 				& $\mathbf{3.22}$ 	& $3.99$			& $10.45$ 			& $4.06$ 			& $3.28$\\
	22 						& - 				& - 				& $4.35$ 			& $1.79$			& $\mathbf{0.89}$ 	& $1.83$ 			& $1.55$\\
	23 						& - 				& - 				& $0.75$ 			& $0.75$			& $\mathbf{0.70}$ 	& $0.82$ 			& $0.82$\\
		\hline
	\textbf{Mean 11}  	& - 				& -  				& $\mathbf{3.61}$ 	& $\mathbf{4.22}$ 	& $\mathbf{5.63}$ 	& $\mathbf{3.35}$ 	& $\mathbf{3.04}$ \\
	\textbf{Std 11}  		& - 				& -  				& $\mathbf{--}$ 	& $\mathbf{9.91}$ 	& $\mathbf{12.98}$ 	& $\mathbf{5.10}$ 	& $\mathbf{4.48}$ \\
	\hline
	\textbf{Mean All}  		& - 				& -  				& $\mathbf{2.28}$ 	& $\mathbf{2.73}$ 	& $\mathbf{3.26}$ 	& $\mathbf{2.30}$ 	& $\mathbf{2.09}$ \\
	\textbf{Std All}  		& - 				& -  				& $\mathbf{3.33}$ 	& $\mathbf{4.01}$ 	& $\mathbf{4.25}$ 	& $\mathbf{3.81}$ 	& $\mathbf{3.42}$ \\
	\hline
	\end{tabular}
\end{table*}

\subsection{Performance Measurement}

The evaluation of performance is carried out in terms of few different performance indices, i.e. average absolute error (AAE), pearson correlation ($r$) and Bland-Altman plot \cite{bland1995comparing}.

For each data recording the AAE is defined as 
\begin{equation}
AAE = \frac{1}{N} \displaystyle\sum\limits_{i=1}^N | B_{est}(i)-B_{true}(i) |,
\end{equation}
where $B_{est}(i)$ and $B_{true}(i)$, respectively, are estimated and ground truth value of heart rate in BPM of the $i$th frame and $N$ is the total number of time frames. Pearson correlation is a measure of degree of similarity between true and estimated values of heart rate. Higher the value of $r$ better the estimates. The Bland-Altman plot measures the agreement between true and estimates of heart rate. Here limit of agreement (LOA) is computed using the average difference ($\mu$) and the standard deviation ($\sigma$), which is defined as $[\mu - 1.96\sigma, \mu + 1.96\sigma]$. LOA is a measurement which shows that $95\%$ data exist within $1.96\sigma$ from mean.

Development of the method and the evaluation are performed in MATLAB 2012b with a 2.2 GHz Intel Processor and 4 GB of RAM.

\section{Results}

Few recently reported heart rate estimation methods have been taken into consideration for the purpose of performance comparison with the proposed method. The methods are TROIKA~\cite{zhang2015troika}, MISPT~\cite{lakshminarasimha2015multiple}, {WFPV}~\cite{temko2015estimation}, TFD~\cite{islam2017cascade}, and CPC~\cite{islam2017time}. Moreover, in order to show the effectiveness of the proposed modified spectral subtraction (referred to as SPECMAR), the algorithm's performance with direct spectral subtraction (referred to as SPECMARWS) is also compared. In both cases, proposed SPECMARWS (using~(\ref{Eq:SS_normal})) and proposed SPECMAR (using~(\ref{Eq:SS_prop})) estimated noise spectrum is obtained using proposed motion artifacts spectrum generation in~(\ref{Eq:ACC}). Table~\ref{Tab:AAE} shows the average absolute error on $23$ recordings when implementing different methods. {The summary of the first $12$ recordings are denoted as \textbf{Mean 12} and \textbf{Std 12} for mean of AAE and standard deviation of absolute errors, respectively. For the rest of the $11$ recordings the summary of metrics are denoted as \textbf{Mean 11} and \textbf{Std 11}. And finally, for all $23$ recordings of the database, the mean of AAE and standard deviation of absolute errors are denoted as \textbf{Mean All} and \textbf{Std All}, respectively.} It is to be noted that in case of TROIKA method, for performance evaluation, some initial time frames are excluded. However, in the proposed method all the time frames are included and none are excluded. It can be observed that, the proposed method provides a better AAE in $5$ of the $23$ recordings and overall mean AAE is lowest among the compared method. 

\begin{figure*}[t]
  \centering
  \includegraphics[scale=0.6]{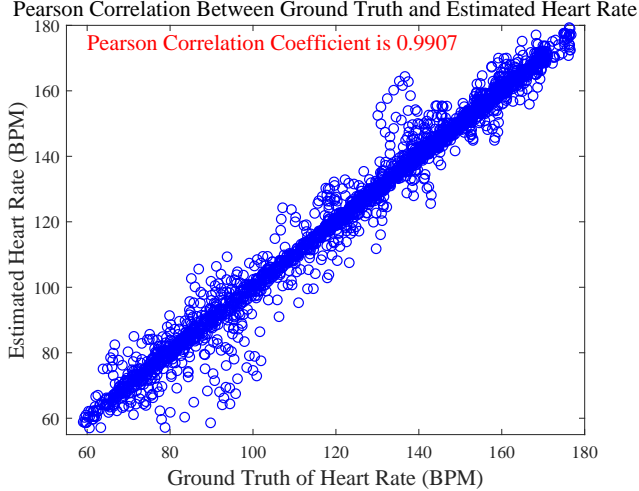}
  \caption{The Pearson correlation of the estimation results on the $23$ data recordings is $0.9907$. On the $12$ recordings pearson correlation is $0.9952$}\label{Fig:pearson}
\end{figure*}

\begin{figure*}[t]
  \centering
  \includegraphics[scale=0.6]{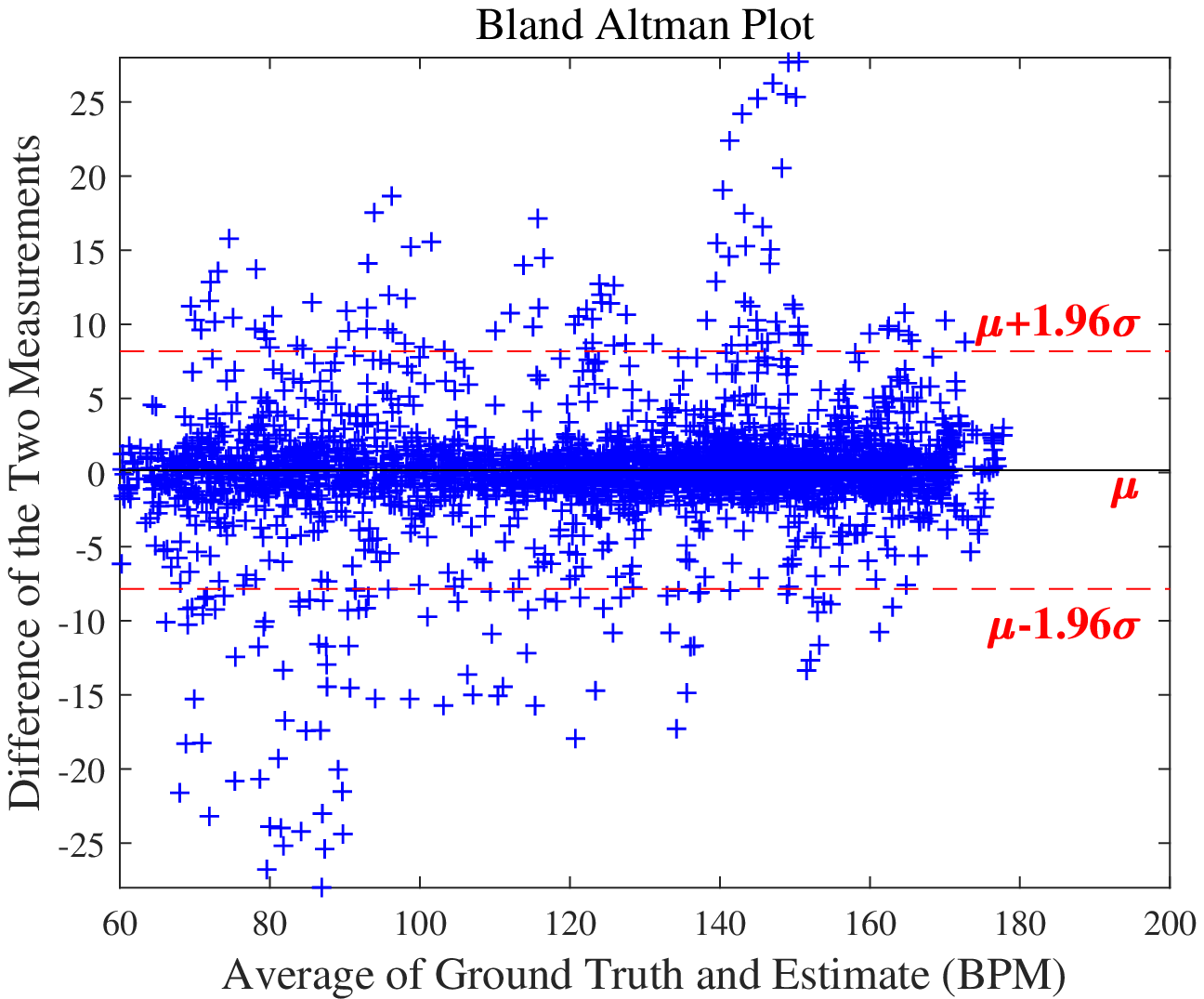}
  \caption{The Bland-Altman plot of the estimation on the 23 recordings.
The LOA is $[-7.85,8.18]$. On first $12$ recordings the LOA is $[-4.78, 4.86]$ BPM.}\label{Fig:bland}
\end{figure*}

The Pearson correlation between ground truth and estimated heart rate is computed and shown in Fig.~\ref{Fig:pearson}. The value of the Pearson coefficient is found $0.9907$ for $23$ recordings and $0.9952$ for $12$ recordings. It is observed from the figure that an approximate linear curve passes close to the origin and that a linear relation exists between ground truth and estimated heart rate. Next, using all the time frames of $23$ recodings, a Bland-Altman plot is computed and shown in Fig.~\ref{Fig:bland}. It is found that limit of agreement is $[-7.85,8.18]$. Furthermore, in the first $12$ recordings the limit of agreement is $[-4.78, 4.86]$.

\begin{figure*}[t]
  \centering
  \includegraphics[scale=0.6]{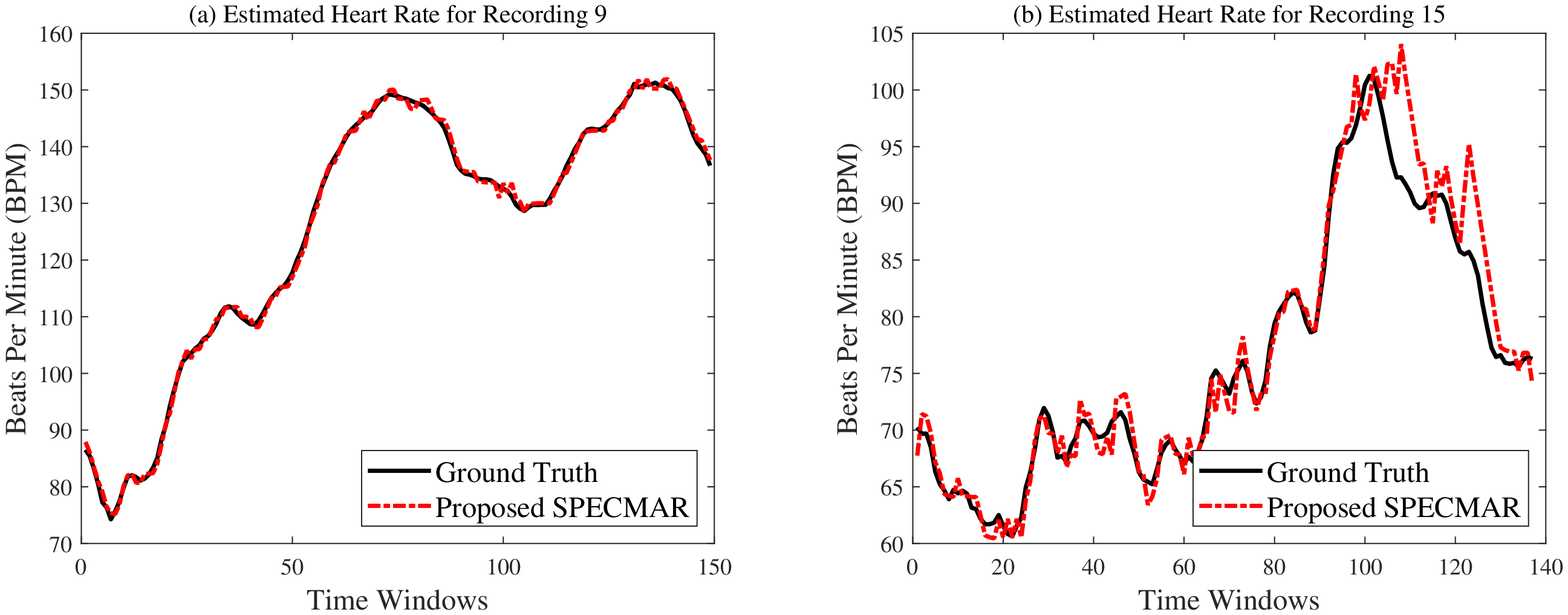}\\
  \caption{{Tracking performance of the proposed SPECMAR for recording (a) $9$ (AAE: $0.64$) and (b) $15$ (AAE: $1.76$).} The estimated heart rate tracks the ground truth heart rate quite well.}\label{Fig:comp}
\end{figure*}

In Fig.~\ref{Fig:comp} estimated BPMs for recording $9$ {and $15$ are} shown. In {the case of recording $9$}, the proposed SPECMAR method performs best and the obtained curve follows the ground truth quite well. {In the case of recording $15$, the proposed SPECMAR provides lowest AAE and can track heart rate fairly well. The method looses tracking around 100th window, however, it gets back on track after few time windows.} It is observed that, when the heart rate changes rapidly along with strong arm movement the spectral signature is severely affected and estimation performance deteriorates.

\begin{figure*}[t]
  \centering
  \includegraphics[scale=0.6]{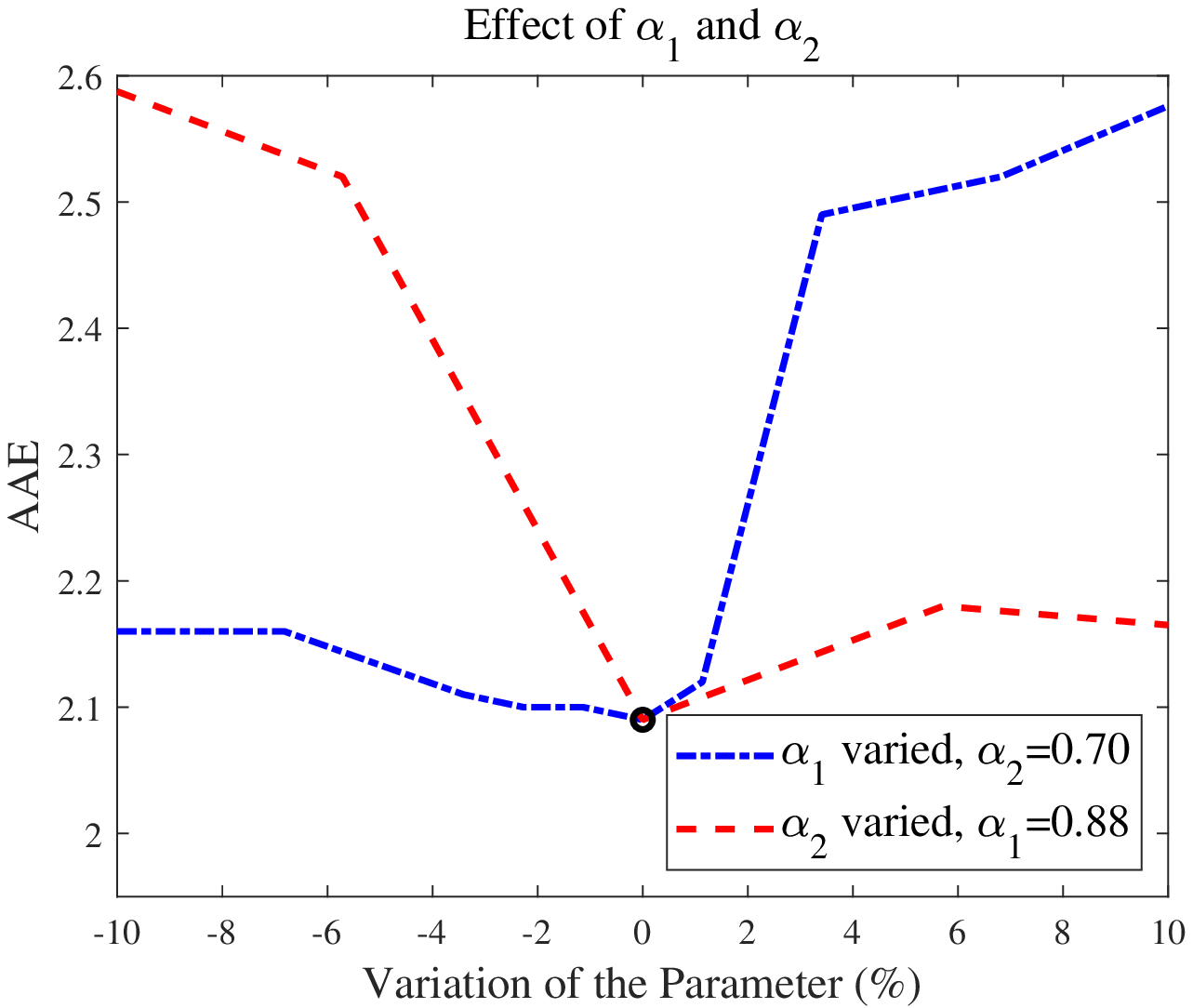}\\
  \caption{{Effet of values of $\alpha_1$ and $\alpha_2$ on AEE. One of the values has been varied keeping the other one fixed. The lowest AAE has been marked using a black circle.}}\label{Fig:variation}
\end{figure*}

{In order to investigate whether the choice of $\alpha_1$ is optimal, the value of $\alpha_1$ has been varied by $\pm10\%$ from the nominal value of $0.88$ while setting the value of $\alpha_2$ to $0.7$  and the mean AAE of 23 recordings have been observed. In a similar way, the value of $\alpha_2$ has been varied from its nominal value of $0.7$ whereas the value of $\alpha_1$ has been set to $0.88$. It is seen from Fig.{~\ref{Fig:variation}} that as the parameters are varied, the AAE first decreases and then increases again. The values $\alpha_1=0.88$ and $\alpha_2=0.70$ provides the lowest AEE which has been marked using a black circle. Additionally, a similar experiment has been conducted to investigate the effect of the number of frequency bins in the spectra $N_{FFT}$. In this experiment, the $N_{FFT}$ has been set to $1024$, $2048$, $4096$, $6144$ and $8192$ and the AAE for $23$ recording are found to be $2.96$, $2.65$, $2.09$, $2.36$ and $2.41$, respectively. The fact that, the value $4096$ provides the lowest AAE is consistent with the results obtained in{~\cite{zhang2015troika}}. Note that, the number of frequency bins in the search region has been scaled according to the value of $N_{FFT}$ in the experiments.}

Finally, the computation time required by the proposed method is compared with those required by some state of the art methods. In Table~\ref{Tab:runtime}, mean runtime for TROIKA, MISPT, {WFPV}, TFD, CPC and the proposed SPECMAR methods are provided. It can be observed from the table that, the proposed method has lowest runtime compared to any of the methods in the tables.

\begin{table}[t]
\centering
\caption{Mean runtime of different methods for the 12 recordings.} \label{Tab:runtime}
\begin{tabular}{c | c }
	\hline
	Name of the Method & Mean Runtime (s)\\
	\hline
    TROIKA~\cite{zhang2015troika} 				& $138.03$\\
	MISPT~\cite{lakshminarasimha2015multiple}	& $1.03$\\
    {WFPV}~\cite{temko2015estimation} 	 	& $0.43$\\
    TFD~\cite{islam2017time}					& $23.03$\\
    CPC~\cite{islam2017cascade}				& $15.11$\\
    SPECMAR 									& $0.24$\\
	\hline
\end{tabular}
\end{table}

\section{Discussions}

In this paper, an efficient method, for motion artifacts removal from PPG signal based on modified spectral subtraction and composite motion artifacts reference generation is proposed.  It has been shown in Figs. \ref{Fig:case1}, \ref{Fig:case2} and \ref{Fig:case3} that only using a single accelerometer channel may not be beneficial depending on the information the channel provides. {Thus, construction of a composite motion artifacts reference has been explored.} It has been observed that the motion artifacts frequencies are generally present in all three axes of the accelerometer albeit having different amplitudes. As a result, motion artifacts can be captured efficiently from the accelerometer signals. It is found that the minimum magnitude of the three accelerometer spectra for a frequency bin efficiently captures the motion artifacts. {It is shown in Figs. {\ref{Fig:ACC_Spectrum}} and {\ref{Fig:ACC_Spectrum_max}} that the proposed composite motion artifacts reference generation method can efficiently capture the motion artifacts.} It has also been stated that the captured PPG signal contains a scaled version of both the noise-free PPG signal and motion artifacts. With this view, the spectral subtraction algorithm has been modified to be a scaled spectral subtraction. {It is shown in Fig.{~\ref{Fig:MSS1}} that the proposed modified spectral subtraction can improve the spectrum quality and facilitate the extraction of true heart rate.} 

In the proposed SPECMAR method, the motion artifacts generation and modified spectral subtraction have been combined. It can be observed from Table~\ref{Tab:AAE} that SPECMAR has lower AAE than SPECMARWS which uses spectral subtraction without scaling. {The high value of Pearson correlation of $0.9907$ and relatively low value of the LOA of $[-7.58,8.18]$ shows that the method can track the heart rate satisfactorily}. Moreover, in order to reduce computational complexity, only the relevant frequency bins (corresponding to 0 to 3.5Hz) have been used in the computation. The overall computation pipeline includes only few filtering operations, spectral subtraction, and a decision mechanism for estimating and tracking the heart rate. As a result, the proposed SPECMAR is faster than many recently published methods as shown in Table~\ref{Tab:runtime}. 

{With respect to previous methods the proposed SPECMAR provides several significant improvements. Similar to the proposed method, the methods, namely, TROIKA, WFPV and TFD employ frequency domain noise removal. TROIKA and TFD use the singular spectrum analysis{~\cite{golyandina2001analysis}} to decompose the PPG signal. Then each of the decomposed signals are processed in spectral domain where the spectral peaks of the decomposed signals are matched with the spectral peaks of the acceleration signal in order to discard some of the decomposed signals. Such a process requires large number of computations for each of the frames. Moreover, the TROIKA method employs computationally expensive MFOCUSS algorithm to construct the spectra of the signals which is the primary cause of the long runtime of the algorithm. The WFPV method performs the simplest Wiener adaptive filtering in the frequency domain to remove motion artifacts from the PPG signals and the phase vocoder part requires extensive time-frequency analysis. The methods, namely, MISPT, TFD and CPC employ time domain noise removal using adaptive filters. The time domain adaptive filter generally suffer from slower convergence rate compared to its frequency domain counterparts and the output has to be computed for each time steps which causes the methods to require extensive computation. In addition, MISPT algorithm requires the spectra of all the previous time steps for computing trajectory and thus causes large memory requirement. On the other hand, TFD method requires cascaded adaptive filtering for noise removal where three adaptive filters are processed serially and the CPC method requires two of such cascaded adaptive filters. As a result, the number of computation is high in these algorithms. On the other hand, the proposed SPECMAR is simplest using only modified spectrum subtraction as the noise removal part and overall employs fewest computations. The computations are further reduced by discarding the frequency bins which are outside the region of interest (greater than $3.5$ Hz). Moreover, unlike the previous methods where both time and frequency domain computation are required in different stages, in the proposed method, the computations are performed in the frequency domain alone.}

\section{Conclusion}

In this paper, an efficient method, for motion artifacts removal from PPG signal based on modified spectral subtraction and composite motion artifacts reference generation is proposed. The motion artifact is captured by taking minimum {amplitude} of each of the frequency bins of the three channel accelerometer. By using this composite motion artifacts in the modified spectral subtraction algorithm, significant noise reduction is achieved which helps in obtaining accurate heart rate estimation. Finally, an efficient tracking scheme is provided to track the heart rate computed from the noise removed spectrum. From extensive experiments on recorded PPG data, it is shown that the proposed SPECMAR method is capable of tracking the ground truth with high estimation accuracy and comparable to recently published methods. Also the method is quantitatively faster than these methods. As a result, it can be used in real-time application without sacrificing estimation accuracy. {The proposed spectral subtraction method can be employed in other areas such as noise removal in speech processing. An improvement of the modified spectral subtraction can be obtained by studying different techniques for setting the parameters $\alpha_1$ and $\alpha_2$ automatically. Additionally, in future alternative tracking algorithm such as the adaptive notch filter can be explored for tracking the heart rates in the algorithm.}

\section*{Acknowledgement}
Authors would like to thank Andriy Temko, Research Fellow, Irish Centre for Fetal and Neonatal Translational Research, Department of EEE, University College Cork for his correspondence regarding computation time of his method. {The authors would also like to thank the anonymous reviewers for their valuable comments that were useful to improve the quality of the paper.}



\begin{thebibliography}{10}
\providecommand{\url}[1]{{#1}}
\providecommand{\urlprefix}{URL }
\expandafter\ifx\csname urlstyle\endcsname\relax
  \providecommand{\doi}[1]{DOI~\discretionary{}{}{}#1}\else
  \providecommand{\doi}{DOI~\discretionary{}{}{}\begingroup
  \urlstyle{rm}\Url}\fi

\bibitem{allen2007photoplethysmography}
Allen, J.: Photoplethysmography and its application in clinical physiological
  measurement.
\newblock Physiological measurement \textbf{28}(3), R1 (2007)

\bibitem{bland1995comparing}
Bland, J.M., Altman, D.G.: Comparing methods of measurement: why plotting
  difference against standard method is misleading.
\newblock The lancet \textbf{346}(8982), 1085--1087 (1995)

\bibitem{fu2008heart}
Fu, T.H., Liu, S.H., Tang, K.T.: Heart rate extraction from photoplethysmogram
  waveform using wavelet multi-resolution analysis.
\newblock J. Medical and Biological Engineering \textbf{28}(4), 229--232 (2008)

\bibitem{gambarotta2016review}
Gambarotta, N., Aletti, F., Baselli, G., Ferrario, M.: A review of methods for
  the signal quality assessment to improve reliability of heart rate and blood
  pressures derived parameters.
\newblock Medical \& biological engineering \& computing \textbf{54}(7),
  1025--1035 (2016)

\bibitem{golyandina2001analysis}
Golyandina, N., Nekrutkin, V., Zhigljavsky, A.A.: Analysis of time series
  structure: SSA and related techniques.
\newblock Chapman and Hall/CRC (2001)

\bibitem{islam2017cascade}
Islam, M.T., Ahmed, S.T., Zabir, I., Shahnaz, C., Fattah, S.A.: Cascade and
  parallel combination ({CPC}) of adaptive filters for estimating heart rate
  during intensive physical exercise from photoplethysmographic signal.
\newblock Healthcare Technology Letters \textbf{5}(1), 18--24 (2018)

\bibitem{islam2017time}
Islam, M.T., Zabir, I., Ahamed, S.T., Yasar, M.T., Shahnaz, C., Fattah, S.A.: A
  time-frequency domain approach of heart rate estimation from
  photoplethysmographic ({PPG}) signal.
\newblock Biomedical Signal Processing and Control \textbf{36}, 146--154 (2017)

\bibitem{jain2014heart}
Jain, P.K., Tiwari, A.K.: Heart monitoring systems — a review.
\newblock Computers in Biology and Medicine \textbf{54}, 1--13 (2014)

\bibitem{kim2006motion}
Kim, B.S., Yoo, S.K.: Motion artifact reduction in photoplethysmography using
  independent component analysis.
\newblock IEEE Trans. Biomedical Engineering \textbf{53}(3), 566--568 (2006)

\bibitem{kyriacou2002investigation}
Kyriacou, P., Powell, S., Langford, R., Jones, D.: Investigation of oesophageal
  photoplethysmographic signals and blood oxygen saturation measurements in
  cardiothoracic surgery patients.
\newblock Physiological measurement \textbf{23}(3), 533 (2002)

\bibitem{lakshminarasimha2015multiple}
Lakshminarasimha~Murthy, N.K., Madhusudana, P.C., Suresha, P., Periyasamy, V.,
  Ghosh, P.K.: Multiple spectral peak tracking for heart rate monitoring from
  photoplethysmography signal during intensive physical exercise.
\newblock IEEE Signal Processing Letters \textbf{22}(12), 2391--2395 (2015)

\bibitem{mccombie2006adaptive}
McCombie, D.B., Reisner, A.T., Asada, H.H.: Adaptive blood pressure estimation
  from wearable {PPG} sensors using peripheral artery pulse wave velocity
  measurements and multi-channel blind identification of local arterial
  dynamics.
\newblock In: in Proc. 28th Annual Int. Conf. IEEE Engineering in Medicine and
  Biology Society, pp. 3521--3524. IEEE, New York City, NY (2006)

\bibitem{periyasamy2017review}
Periyasamy, V., Pramanik, M., Ghosh, P.K.: Review on heart-rate estimation from
  photoplethysmography and accelerometer signals during physical exercise.
\newblock J. Indian Institute of Science pp. 1--12 (2017)

\bibitem{seyedtabaii2008kalman}
Seyedtabaii, S., Seyedtabaii, L.: Kalman filter based adaptive reduction of
  motion artifact from photoplethysmographic signal.
\newblock World Academy of Science, Engineering and Technology \textbf{37},
  173--176 (2008)

\bibitem{tamura2018photoplethysmogram}
Tamura, T., Maeda, Y.: Photoplethysmogram.
\newblock In: Seamless Healthcare Monitoring, pp. 159--192. Springer (2018)

\bibitem{tamura2014wearable}
Tamura, T., Maeda, Y., Sekine, M., Yoshida, M.: Wearable photoplethysmographic
  sensors—past and present.
\newblock Electronics \textbf{3}(2), 282--302 (2014)

\bibitem{temko2015estimation}
Temko, A.: Estimation of heart rate from photoplethysmography during physical
  exercise using wiener filtering and the phase vocoder.
\newblock In: 37th Annual International Conference of the IEEE Engineering in
  Medicine and Biology Society, pp. 1500--1503. IEEE, Milano, Italy (2015)

\bibitem{yousefi2014motion}
Yousefi, R., Nourani, M., Ostadabbas, S., Panahi, I.: A motion-tolerant
  adaptive algorithm for wearable photoplethysmographic biosensors.
\newblock IEEE J. Biomedical and Health Informatics \textbf{18}(2), 670--681
  (2014)

\bibitem{zhang2015photoplethysmography}
Zhang, Z.: Photoplethysmography-based heart rate monitoring in physical
  activities via joint sparse spectrum reconstruction.
\newblock IEEE Trans. Biomed. Eng. \textbf{62}(8), 1902--1910 (2015)

\bibitem{zhang2015troika}
Zhang, Z., Pi, Z., Liu, B.: {TROIKA}: A general framework for heart rate
  monitoring using wrist-type photoplethysmographic signals during intensive
  physical exercise.
\newblock Biomedical Engineering, IEEE Transactions on \textbf{62}(2), 522--531
  (2015)

\end{thebibliography}

\end{document}